\documentclass[twocolumn]{aastex63}

\usepackage{CJKutf8}
\usepackage{hyperref}

\newcommand{\cntext}[1]{\begin{CJK}{UTF8}{gbsn}#1\end{CJK}\kern-1ex}
\newcommand{\noop}[1]{}

\received{2020 March 17}
\revised{2020 April 29}
\accepted{2020 May 4}
\turnoffedit

\shorttitle{Microwave imaging spectroscopy of an erupting flux rope}
\shortauthors{Chen et al.}
\graphicspath{{./}{figures/}}

\begin{document}

\title{Microwave Spectral Imaging of an Erupting Magnetic Flux Rope: Implications for the\\ Standard Solar Flare Model in Three Dimensions}

\correspondingauthor{Bin Chen}
\email{bin.chen@njit.edu}

\author[0000-0002-0660-3350]{Bin Chen (\cntext{陈彬})}
\affiliation{Center for Solar-Terrestrial Research, New Jersey Institute of Technology, 323 M L King Jr. Blvd., Newark, NJ 07102-1982, USA}

\author[0000-0003-2872-2614]{Sijie Yu (\cntext{余思捷})}
\affiliation{Center for Solar-Terrestrial Research, New Jersey Institute of Technology, 323 M L King Jr. Blvd., Newark, NJ 07102-1982, USA}

\author[0000-0002-6903-6832]{Katharine K. Reeves}
\affiliation{Harvard-Smithsonian Center for Astrophysics, 60 Garden St., Cambridge, MA 02138, USA}

\author[0000-0003-2520-8396]{Dale E. Gary}
\affiliation{Center for Solar-Terrestrial Research, New Jersey Institute of
Technology, 323 M L King Jr. Blvd., Newark, NJ 07102-1982, USA}

\begin{abstract}
We report microwave spectral imaging observations of an erupting magnetic flux rope during the early impulsive phase of the X8.2-class limb flare on 2017 September 10, obtained by the Expanded Owens Valley Solar Array. A few days prior to the eruption, when viewed against the disk, the flux rope appeared as a reverse S-shaped dark filament along the magnetic polarity inversion line. During the eruption, the rope exhibited a ``hot channel'' structure in extreme ultraviolet and soft X-ray passbands sensitive to $\sim$10 MK plasma. The central portion of the flux rope was nearly aligned with the line of sight, which quickly developed into a teardrop-shaped dark cavity during the early phase of the eruption. A long and thin plasma sheet formed below the cavity, interpreted as the reconnection current sheet viewed edge on. A nonthermal microwave source was present at the location of the central current sheet, which extended upward encompassing the dark cavity. A pair of nonthermal microwave sources were observed for several minutes on both sides of the main flaring region. They shared a similar temporal behavior and spectral property to the central microwave source below the cavity, interpreted as the conjugate footpoints of the erupting flux rope. These observations are broadly consistent with the magnetic topology and the associated energy release scenario suggested in the three-dimensional standard model for eruptive solar flares. In particular, our detection of nonthermal emission at conjugate flux rope footpoints provides solid evidence of particle transport along an erupting magnetic flux rope.

\end{abstract}


\section{Introduction} \label{sec:intro}
Magnetic flux ropes are believed to be the centerpiece of the three-part structure \citep{1985JGR....90..275I} of coronal mass ejections (CMEs), which are major drivers for space weather (see, e.g., a review by \citealt{2012LRSP....9....3W}). In the standard solar flare model (or the ``CSHKP'' model, after \citealt{1964NASSP..50..451C, 1966Natur.211..695S, 1974SoPh...34..323H,1976SoPh...50...85K}), the eruption of flux ropes also induces the impulsive flare energy release through magnetic reconnection. Signatures of flare-associated flux rope eruptions in the solar corona have been frequently reported in extreme ultraviolet (EUV) wavelengths, particularly the so-called EUV ``hot blob'' or ``hot channel'' structures (e.g., \citealt{2011ApJ...732L..25C,2013ApJ...763...43C,2011ApJ...727L..52R,2012NatCo...3..747Z,2014ApJ...794..149C,2014ApJ...792L..40S,2015ApJ...808..117N,2019ApJ...871...25W}, and a recent review by \citealt{2017ScChE..60.1383C}).

Recently, based on magnetohydrodynamic (MHD) simulations, the standard model has been extended into three dimensions (3D; \citealt{2012A&A...543A.110A,2013A&A...549A..66A,2013A&A...555A..77J,2014ApJ...788...60J}. The 3D flare model has been able to reproduce observed features such as twisted flux-rope-like structures, S-shaped sigmoids, shape of flare ribbons, and apparent slipping motion of flare arcades \citep[see, e.g.,][for a review]{2015SoPh..290.3425J}. However, the current MHD framework does not include kinetic processes, which are crucial not only for dissipating the magnetic energy, but also for efficiently accelerating particles to high energies. These energetic particles play a key role in the total flare energy budget \citep{2012ApJ...759...71E,2016ApJ...832...27A}. They are also primarily responsible for prompt transport of the released energy throughout the flaring volume, resulting in a variety of flare phenomena (see a review by \citealt{2017LRSP...14....2B}). Further development of the standard model will require the inclusion of kinetic processes (see, e.g., \edit2{new frameworks} proposed by \edit2{\citealt{2018ApJ...866....4L,2019PhPl...26j2903A,2019PhPl...26a2901D}}). Meanwhile, validation of the standard model calls for observational tests based on emission from flare-accelerated nonthermal particles.

Reports of the nonthermal counterpart of the erupting flux ropes are relatively rare in the literature. There have been a few reports on hard X-rays (HXRs; e.g., \citealt{1992ApJ...390..687K,2001ApJ...561L.211H,2007ApJ...669L..49K,2013ApJ...779L..29G}). Nonthermal microwave emission, produced by a non-Maxwellian distribution of electrons gyrating in the coronal magnetic field, offers a unique view of the flux rope field lines rendered visible by the flare-accelerated nonthermal electrons \citep{2014ApJ...787..125N,2016ApJ...820L..29W}. 
When spectrally resolved imaging data are available with adequate bandwidth, spectral sampling, and temporal cadence, they can also be used to constrain the spatial distribution and temporal evolution of the magnetic field and nonthermal electrons (see, e.g., recent studies by \citealt{2018ApJ...863...83G, Chen2020,  Fleishman2020, Kuroda2020}). However, such diagnostics for accelerated electrons and magnetic field of the flux ropes in the low corona have been illusive, mainly due to the lack of microwave imaging spectroscopy observations. We note only a handful of studies that reported radio counterparts of flux ropes/CMEs much higher in the corona, based on low spatial resolution data in decimetric/metric wavelengths \citep[e.g.,][]{2001ApJ...558L..65B,2007ApJ...660..874M,2013ApJ...766..130T,2017A&A...608A.137C, 2019arXiv190912041M}.

Our microwave observations of the erupting flux rope were obtained by the Expanded Owens Valley Solar Array (EOVSA) at 2.5--18 GHz during the early impulsive phase of the X8.2-class flare on 2017 September 10 (SOL2017-09-10). Many aspects of this flare event have already been studied by numerous works, which include the flux rope eruption and the white-light CME \citep{2018ApJ...852L...9S,2018ApJ...868..107V,2018ApJ...853L..18Y}, the large-scale current-sheet-like structure \citep{2018ApJ...868..148L,2018ApJ...852L...9S,2018ApJ...853L..18Y,2019ApJ...887L..34F,Chen2020} with signatures of turbulence \citep{2018ApJ...866...64C,2018ApJ...854..122W,2018ApJ...864...63P}, plasma outflows and quasi-periodic pulsations \citep{2018ApJ...868..148L,2018ApJ...866...64C,2019ApJ...875...33H}, global EUV waves and large-scale shocks \citep{2018ApJ...864L..24L,2019ApJ...878..106H,2019ApJ...871....8L,2019NatAs...3..452M}, nonthermal emissions by flare- and shock-accelerated electrons \citep{2018ApJ...863...83G,2018ApJ...865L...7O,2019NatAs...3..452M,Fleishman2020,2020ApJ...889...72K}, solar energetic particle events \citep{2018ApJ...863L..39G,2019SpWea..17..419B,2020ApJ...890...13K}, and ground level enhancement events detected on Earth \citep{2018SoPh..293..136M,2019SoPh..294...22K} and Mars \citep{2018SpWea..16.1156G}. In particular, an EUV hot channel structure was observed by SDO/AIA 131 \AA\ passbands that went through a slow-rise to fast-eruption phase during the flare impulsive phase, suggested as the main driver for the event \citep{2018ApJ...853L..18Y,2018ApJ...868..107V}. Here we focus on the early impulsive phase of the event when the flux rope could still be observed in the low corona. Complemented by EUV and SXR data, our microwave imaging spectroscopy observations reveal a never-before-seen, detailed picture of the flux rope illuminated by flare-accelerated electrons. These observations also allow us to diagnose the magnetic properties of the flux rope and the accelerated electrons in a broad flare region. We present the main observational results in Section \ref{sec:obs}. In Section \ref{sec:discussion}, we place the results in the context of the 3D standard flare model and discuss their implications in flare energy release, electron acceleration, and electron transport.

\section{Observations} \label{sec:obs}
\subsection{The Preexisting filament}\label{sec:obs:fila}

\begin{figure*}[ht!]
\epsscale{1.2}
\plotone{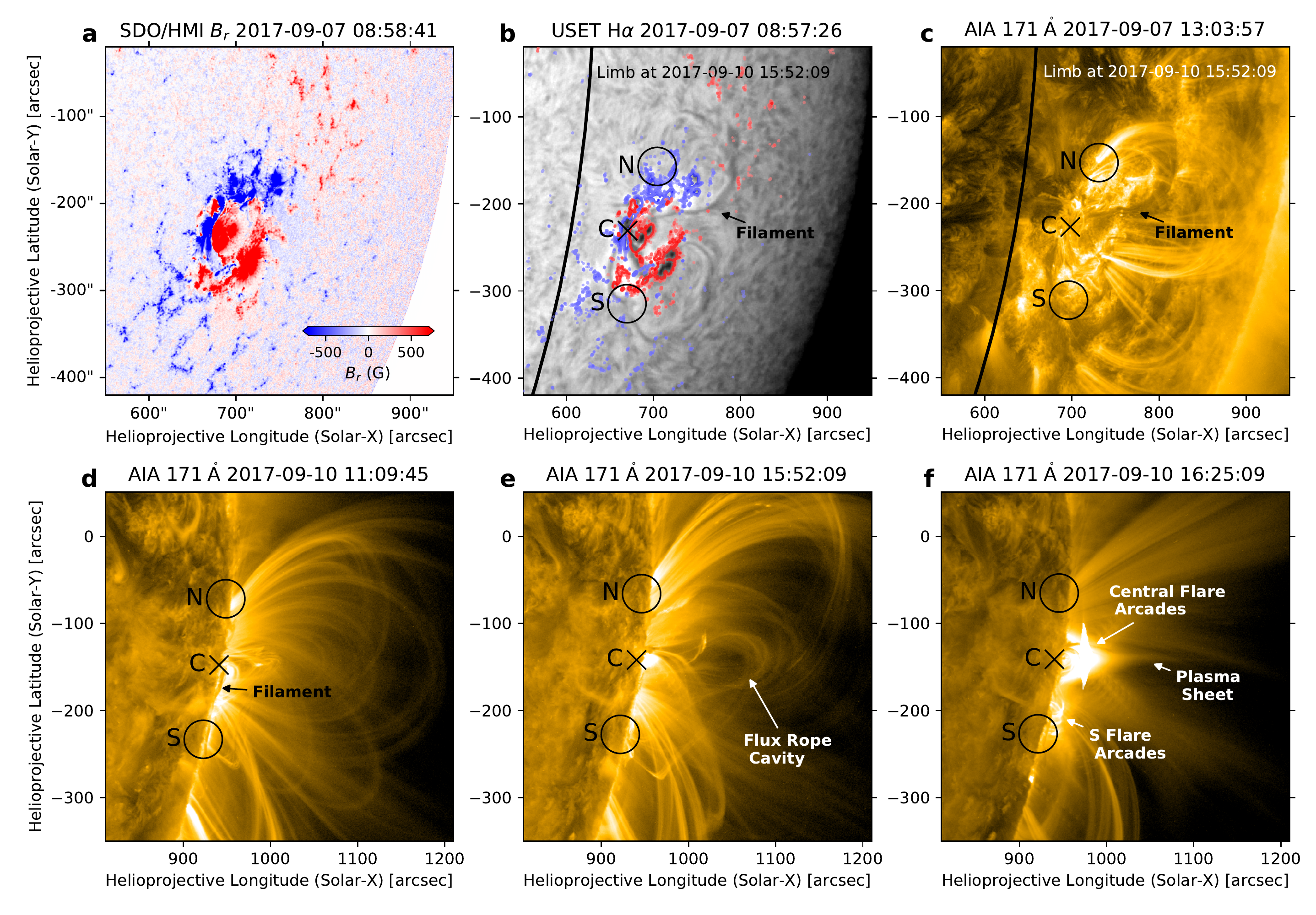}
\caption{Preexisting filament that led to the X8.2 eruptive solar flare event. (a) SDO/HMI radial photospheric magnetogram on 2017 September 7. Red and blue indicate positive and negative magnetic polarity, respectively. (b) H-$\alpha$ image obtained by the USET telescope at a similar time as (a), shown in grayscale, with the magnetogram overlaid. 
The filament is seen as a reverse S-shaped dark structure located close to the magnetic polarity inversion line (PIL). Thick black curve indicates the limb location during the eruption on 2017 September 10. (c)--(f) SDO/AIA 171 \AA\ images of the dark filament seen against the disk (c), $\sim$5 hr prior to the X8.2 event (d), during the eruption (e), and after the flare peak (f). Symbol ``X'' marks the center of the filament. Circles marked with ``N'' and ``S'' denote the northern and southern footpoints of the filament/flux rope, respectively. Their position in each panel is the same after compensation for solar differential rotation. An animation is available at \href{https://web.njit.edu/~binchen/download/publications/Chen+2020_MFR/Figure1_video.mp4}{this link} for the H-$\alpha$ and AIA 171 \AA\ observations from 17:05 UT on 2017 September 7 to 16:42 UT on 2017 September 10. The realtime duration of the video is 19 s.}\label{fig:fila}
\end{figure*}

The X8.2 event under study originated from NOAA Active Region (AR) 12673 when it rotated to the western limb on 2017 September 10. Three days prior to the event when the AR was viewed against the disk, a dark, reverse S-shaped filament was seen in both the H-$\alpha$ 6563 \AA\ and SDO/AIA 171 \AA\ images (Figures \ref{fig:fila}(b) and (c)). By comparing the observed filament location to the radial photospheric magnetic field map ($B_r$; obtained by the Helioseismic and Magnetic Imager (HMI) on board SDO; \citealt{2014SoPh..289.3483H}) shown in Figure \ref{fig:fila}(a), we find that the filament was located along the magnetic polarity inversion line (PIL) that separates the positive and negative polarities of the main sunspot groups (Figure \ref{fig:fila}(b)). The northern and southern ends of the filament were anchored near the edge of the sunspot group with a negative and positive magnetic polarity, respectively (marked in Figures \ref{fig:fila}(b) and (c) as circles). Near the central region of the AR (marked by a symbol ``X'' in Figure \ref{fig:fila}(b)), the reverse S-shaped filament and the PIL display a sharp turn from a general north--south orientation toward the east--west orientation. Despite the occurrence of numerous relatively small flare events between 2017 September 7 and 10, this filament could be clearly distinguished as it rotated to the west limb (see the animation accompanying Figure \ref{fig:fila}). 

When the filament rotated to the west limb on 2017 September 10 (Figure \ref{fig:fila}(d), showing SDO/AIA 171 \AA\ image at 11 UT, $\sim$5 hr prior to the eruption), the northern branch of the filament was completely occulted by the limb (as it was located further west). Only the southern branch of the filament remained visible. The central part of the filament, which initially had an east--west orientation viewed against the disk, was then aligned nearly along the line of sight (LOS; marked by ``X'' in Figure \ref{fig:fila}(d), which is the same location as in earlier times shown in panels (a)--(c) after compensating for solar rotation).
As will be described next, during the impulsive phase of the event this location was directly below the erupting dark cavity (see Figure \ref{fig:fila}(e)). The latter strongly implicates that the erupting flux rope at this central location was viewed along its axis, consistent with the geometry of the filament in the preeruption phase. 

\begin{figure*}[ht!]
\epsscale{1.2}
\plotone{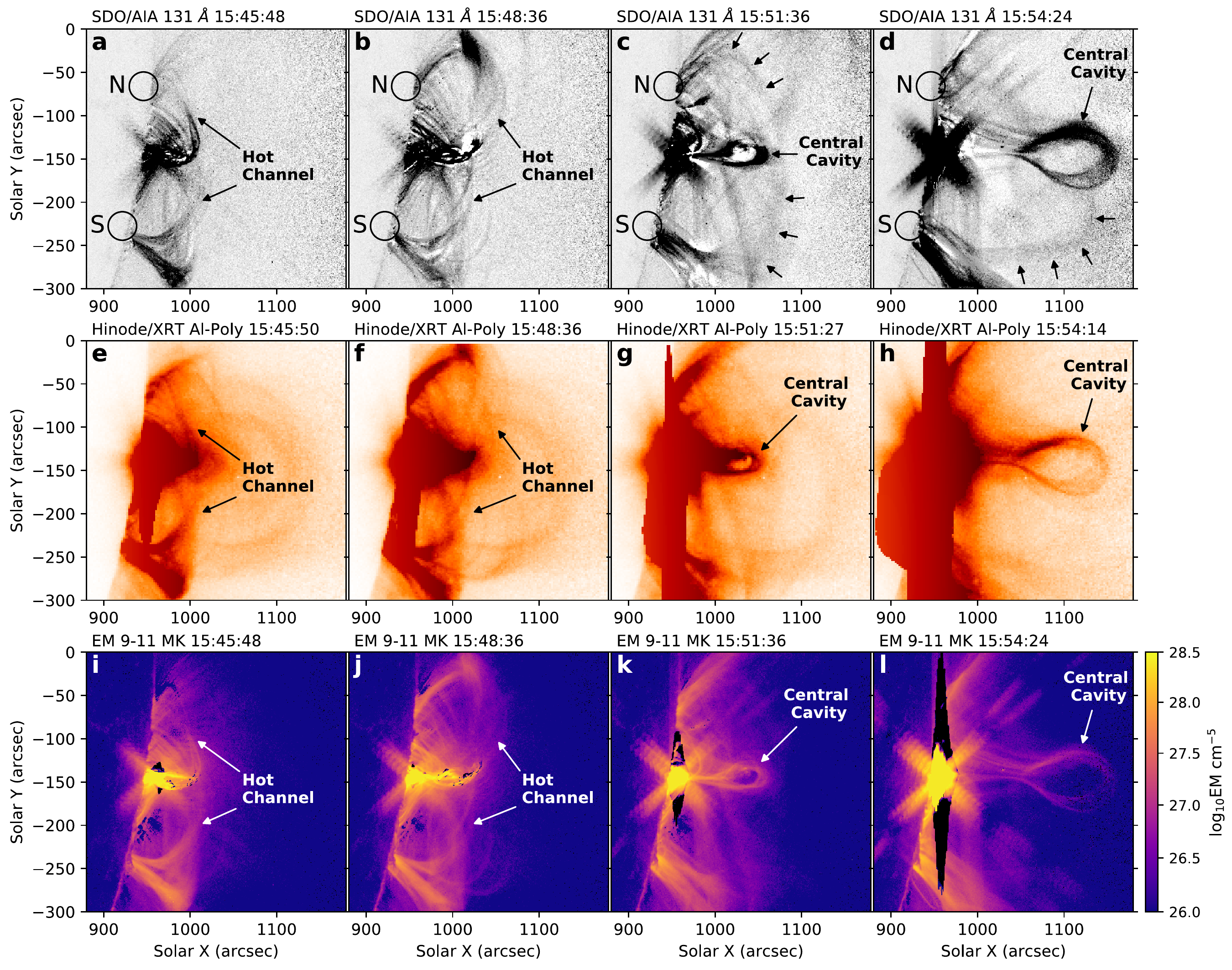}
\caption{Erupting flux rope seen as a hot channel structure with a central dark cavity in EUV and SXR. (a)--(d) Eruption seen in the SDO/AIA EUV 131 \AA\ passband (shown in reverse grayscale) with dynamic features enhanced using the unsharp mask technique. This passband is sensitive to $\sim$10 MK plasma. The hot channel structure features a bright core that later develops into a teardrop-shaped cavity. Multiple strands are present at both sides of the cavity (marked by short arrows), interpreted as the flux rope legs connecting to both the northern and southern footpoints (circles; identified from the preeruption filament shown in Figure \ref{fig:fila}). (e)--(h) The same eruption observed by Hinode/XRT's Al-Poly passband in SXR, which has a peak response of $\sim$8 MK. (i)--(l) Emission measure maps of hot plasma at 9--11 MK, derived from six SDO/AIA EUV passband images at 94, 131, 171, 193, 211, and 335 \AA. \label{fig:fr_euv}}
\end{figure*}

\subsection{Flux Rope Eruption: EUV and SXR Observations}\label{sec:obs:erupt}

On 2017 September 10 at $\sim$15:30 UT, the preexisting filament showed a slow rising motion before it started to accelerate at $\sim$15:46 UT \citep{2018ApJ...853L..18Y}. The acceleration of the filament peaked at $\sim$15:54 UT \citep{2018ApJ...868..107V}. During the early rise phase of the filament, it was visible as a hot channel structure in the SDO/AIA 131 \AA\ passband (Figures \ref{fig:fr_euv}(a)--(d)), which has a response to the \ion{Fe}{21} line sensitive to $\sim$10 MK plasma \citep{2010A&A...521A..21O}. Dynamic features in the 131 \AA\ images are enhanced using an ``unsharp masking'' technique (see, e.g., \citealt{2018ApJ...852L...9S} for details). The hot nature of the AIA 131 \AA\ hot channel structure during the eruption was further confirmed by its SXR counterpart observed by the Soft X-Ray telescope on board Hinode (Hinode/XRT; \citealt{2007SoPh..243...63G}) (Figures \ref{fig:fr_euv}(e)--(h)), as well as the column emission measure (EM; defined as $\xi = n^2_e L$, where $n_e$ is the thermal electron density and $L$ is the column depth; Figures \ref{fig:fr_euv}(i)--(l)) in 9--11 MK derived from six SDO/AIA passband images at 94, 131, 171, 193, 211, and 335 \AA. The differential emission measure (DEM) analysis was carried out using the routine {\tt xrt\_dem\_iterative2} \citep{Golub2004,Weber2004}, which has been extensively tested \citep{Cheng2012}, and provides results similar to other DEM algorithms \citep{SchmelzKashyap2009,SchmelzSaar2009,HannahKontar2012}. The XRT Al-Poly images were not used for the quantitative DEM analysis due to the possible white-light contamination in this filter.

The main axis of the hot channel structure, shown in Figure \ref{fig:fr_euv}, was nearly parallel to the west limb in the north--south direction, consistent with the overall north--south geometry of the reverse S-shaped filament when viewed against the disk (see Figure \ref{fig:fila}(b)). Near the center of the hot channel structure, a bright core was present, connected by hot channel strands from both the northern and southern sides. Soon the central core developed into a teardrop-shaped cavity with a bright rim (right two columns in Figure \ref{fig:fr_euv}; see also \citealt{2018ApJ...868..107V,2018ApJ...853L..18Y}). A long and thin plasma sheet appeared just below the cavity, which connected to the underlying flare arcade with a cusp shape. The plasma sheet, best seen in hot AIA passbands and SXR, is likely heated locally in the corona to $>$10 MK \citep{2018ApJ...866...64C,2018ApJ...854..122W,2018ApJ...853L..18Y}.

The inferred orientation of the flux rope axis at the central location is fully consistent with the geometry of the preexisting filament described in the previous subsection: the central portion of the reverse S-shape filament had an east--west orientation when it was viewed against the disk a few days before (marked with the ``X'' symbol, the same location as in Figure \ref{fig:fr_euv} but adjusted according to solar rotation). When it rotated to the limb, its axis at the central location was thus nearly aligned with the LOS direction. The northern and southern footpoints of the filament, when viewed against the disk (marked in Figure \ref{fig:fila} as circles), are also consistent with the footpoints of the hot channel structure during the eruption (Figure \ref{fig:fr_euv}). 

\subsection{Flux Rope Eruption: Microwave Imaging Spectroscopy}\label{sec:obs:eovsa}

EOVSA observed the Sun from $\sim$14:30 UT to 01:10 UT of the next day for more than 10 hr. It had full coverage of the X8.2 event from its onset at $\sim$15:35 UT well into the decay phase. EOVSA obtained data in 2.5--18 GHz with 134 frequency channels spread over 31 equally spaced spectral windows (``SPWs''), each of which has a bandwidth of 160 MHz. The center frequencies of these SPWs are given by $\nu=2.92 + n/2$ GHz, where $n$ is the SPW number from 0 to 30. An overview of the EOVSA observations of this event in different flare phases and initial imaging spectroscopy results were discussed in \citet{2018ApJ...863...83G}. More in-depth studies that utilize the newly available diagnostics for the coronal magnetic field and energetic electrons were reported in \citet{Fleishman2020} and \citet{Chen2020}.

This study uses the same dataset reported in \citet{2018ApJ...863...83G} and \citet{Chen2020} during the early impulsive phase of the event. In this work, we further refined the calibration and self-calibration procedures, and adopted a multifrequency synthesis (MFS) imaging technique to increase the dynamic range of the microwave images. This practice helped to reveal greater details of the microwave counterpart of the flux rope. We also recovered imaging at the lowest-frequency band (SPW \#0; centered at 2.9 GHz) using self-calibration, which lacked calibrations against a celestial source and was not analyzed in our earlier works. This band, as we will discuss later, shows a new feature that is likely associated with the extension of the central current sheet toward north and south with a face-on viewing geometry. Details about the self-calibration and MFS imaging techniques are discussed in the Appendix \ref{ap:imaging}.

\begin{figure*}[ht!]
\epsscale{1.1}
\plotone{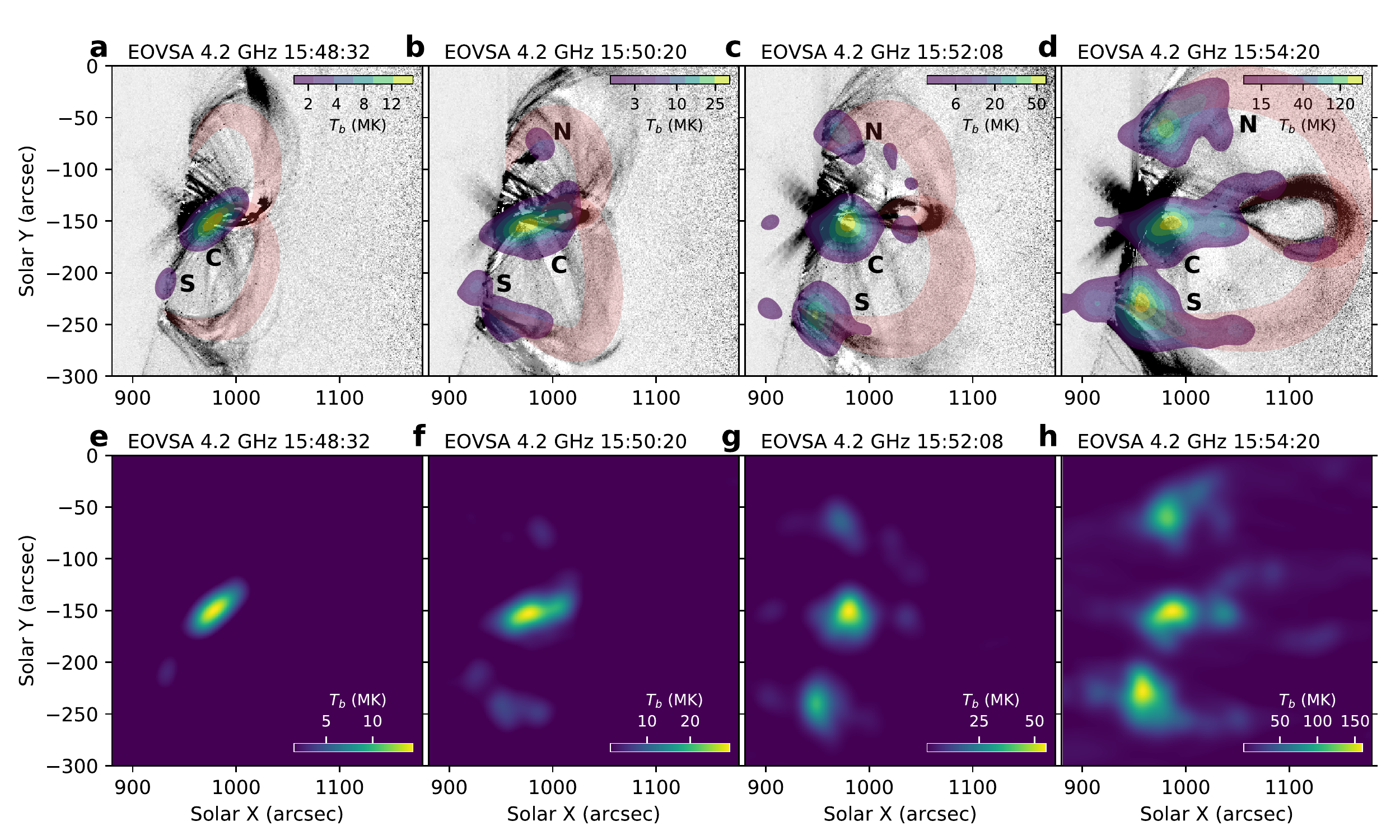}
\caption{Microwave imaging of the erupting flux rope. (a)--(d) SDO/AIA 131 \AA\ images at four times during the eruption, enhanced using the unsharp masking technique (similar to Figure \ref{fig:fr_euv}(a)--(d)). EOVSA contours from panels (e)--(h) are superposed. The flux rope structure seen in 131 \AA\ (see, Figure \ref{fig:fr_euv}(a)--(d)) is outlined in pink. (e--h) EOVSA images at the same four times, made with multifrequency synthesis imaging using 11 spectral channels spanning the 3.85--4.50 GHz frequency range (centered at 4.2 GHz). Colors in each panel represent brightness temperature ($T_b$), linearly scaled to its maximum value as shown in the inset colorbars. \label{fig:fr_mw}}
\end{figure*}

\begin{figure*}[ht!]
\epsscale{1.1}
\plotone{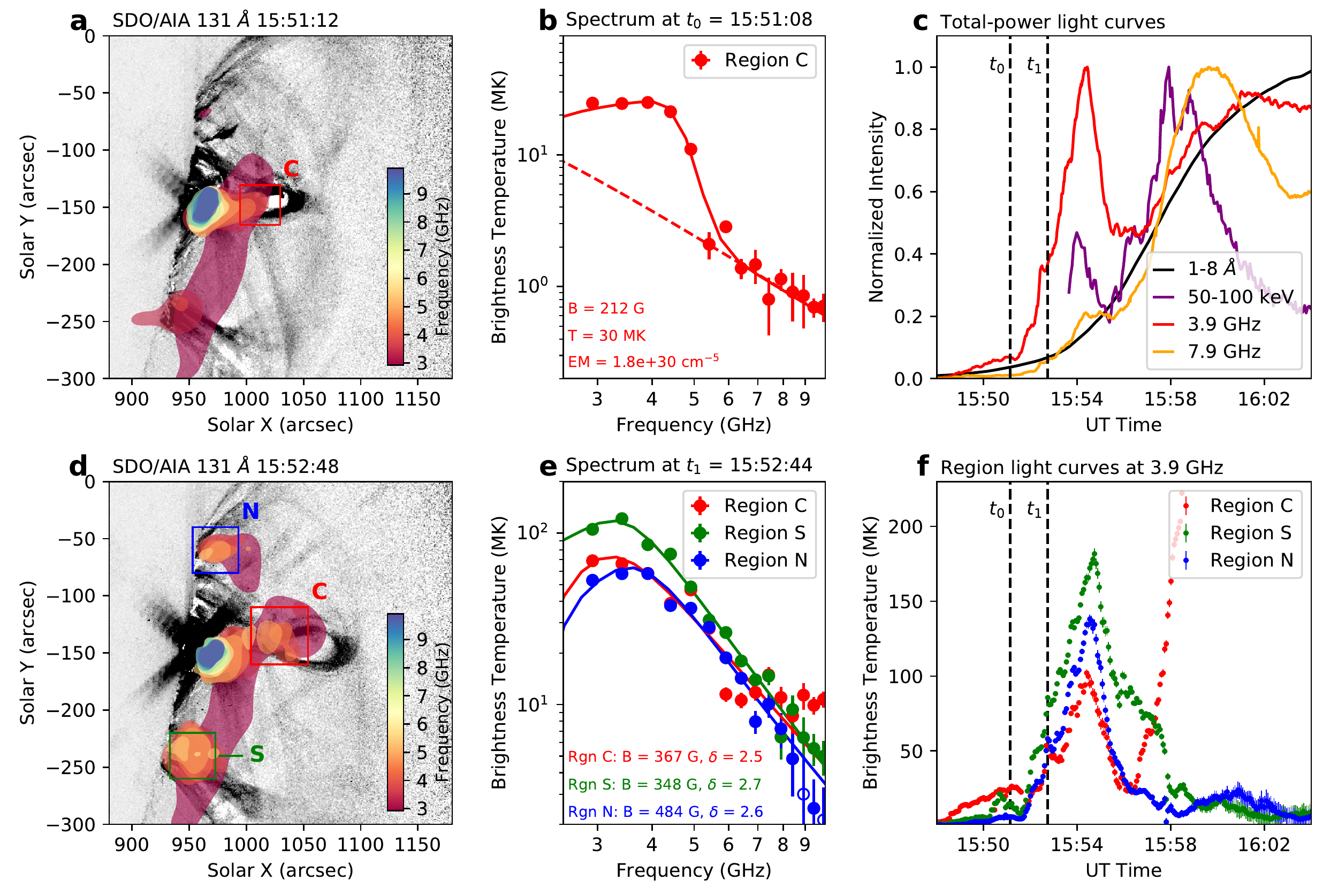}
\caption{Spatially resolved microwave spectra and light curves of the erupting flux rope. (a) and (d) EOVSA multifrequency microwave images at two selected times ($t_0 = $15:51:08 UT and $t_1=$15:52:44 UT). Filled contours from red to blue indicate increasing microwave frequency (showing 32\% of the maximum brightness at each frequency). Background images are SDO/AIA 131 \AA\ images at the closest time, enhanced using the unsharp masking technique (shown in inverse color scale---black means greater intensity). (b) Microwave brightness temperature spectrum $T_b(\nu)$ obtained from the flux rope core region (region ``C,'' marked as a red box in (a)) at $t_0 = $15:51:08 UT. Each value at a given frequency represents the peak $T_b$ within the selected region. Solid curve shows the best-fit results based on thermal gyrosynchrotron and free-free emission. Dashed line shows the free-free component that dominates the high-frequency portion of the spectrum. (e) Similar to (b), but the microwave brightness temperature spectra are for $t_1 = $15:52:44 UT obtained from three different regions marked in (d), which correspond to the flux rope core (red box ``C''), northern flux rope footpoint (blue box ``N''), and southern flux rope footpoint (green box ``S''), respectively. Solid curves are best-fit results based on nonthermal gyrosynchrotron emission. (c) Total-power (i.e., full-Sun integrated) GOES SXR 1--8 \AA (black), EOVSA microwave 3.9 GHz (red), 7.9 GHz (orange), and RHESSI HXR 50--100 keV (purple) light curves. The two selected times are indicated by the vertical dashed lines. (f) Spatially resolved microwave brightness temperature light curves at 3.9 GHz, obtained from the three selected regions in (d). They show similar temporal evolution during the first impulsive peak around 15:54 UT. The steep increase of the central source ``C'' (red curve) after $\sim$15:58 UT is due to the rising microwave source at the looptop entering the selected region (red box) during the main flare peak. \label{fig:spec}}
\end{figure*}

By combining 11 spectral channels spanning 3.85--4.50 GHz (centered at 4.2 GHz), the MFS-enhanced microwave images showed a striking similarity in morphology and evolution between the microwave and EUV 131 \AA\ observations of the erupting flux rope (Figure \ref{fig:fr_mw}). At the very beginning of the eruption (Figure \ref{fig:fr_mw}(a) and (e)), the central flux rope core can be clearly distinguished in microwaves as a compact source (marked as ``C'' in Figure \ref{fig:fr_mw}). A faint microwave source near the southern footpoint of the flux rope was also visible (marked as ``S''), which then gradually developed into an elongated source (Figure \ref{fig:fr_mw}(b) and (f)). The upper tip of the elongated southern microwave source aligned very well with the southern leg of the 131 \AA\ hot channel structure (Figure \ref{fig:fr_mw}(b)--(d)). The microwave counterpart of the northern footpoint of the flux rope appeared later (from $\sim$15:50 UT; marked as ``N''). This delay is likely a result of the northern portion of the flux rope being located further west and therefore being occulted much more than its southern counterpart (see, Figures \ref{fig:fila}(b) and (c)). As the flux rope rose to higher heights and the accompanying flare emission further increased, the gradually brighter microwave sources displayed more complexity in their morphology. In particular, as shown in Figure \ref{fig:fr_mw}(c) and (g), a number of discrete microwave sources were visible northward of the flux rope core and appeared to align along the northern portion of the SDO/AIA 131 \AA\ hot channel structure.

The most striking spatial correspondence between the 4.2 GHz microwave source and the EUV/SXR hot channel structure was seen around 15:54 UT during the first impulsive HXR/microwave peak of the flare, shown in Figures \ref{fig:fr_mw}(d) and (h). At that time, in EUV 131 \AA\ and SXR images, the flux rope core developed fully into a teardrop-shaped dark cavity. Similar to the 131 \AA\ image, the concurrent microwave image shows a faint cavity-shaped structure encompassing the EUV/SXR cavity. Moreover, both the northern and southern microwave sources show a clear extension from the flux rope footpoints toward higher heights, which also appears to follow the 131 \AA\ hot-channel strands connecting to the cavity.

At EOVSA's lowest-frequency band (SPW \#0, centered at 2.9 GHz), the microwave source appears to connect the central and southern source (red contours in Figures \ref{fig:spec}(a) and (d)). The elongated source is oriented nearly parallel to the limb, similar to the orientation of the flux rope axis southward from the core. It appears to be located at the top of the series of post-flare arcades southward of the main flaring region, which also had the same north--south orientation (Figure \ref{fig:fila}(f)) but become visible in EUV only later. Although this elongated looptop source is nearly invisible at higher microwave frequencies, it has a peak brightness temperature exceeding 100 MK after 15:52:40 UT, therefore it is likely nonthermal. From the geometry we believe the 2.9~GHz source lies parallel to but below the flux rope, in the region of a face-on reconnecting current sheet that is busy forming the southern arcade (see further discussion in Section~\ref{sec:discussion}).

To better investigate the nature of the microwave sources, we produce a time series of multiband microwave images at all available 31 SPWs from 2.9 to 18 GHz with 4 s cadence. To increase the image dynamic range, frequency channels within each of the 31 SPWs are combined to form a single image for each band. These 31-band images allow us to derive spatially resolved microwave brightness temperature spectra $T_b(\nu)$ from any selected spatial regions of interest. In Figures~\ref{fig:spec}(b) and (e), we show the microwave spectra derived from selected regions marked in Figure~\ref{fig:spec}(a) and (d) at 15:51:08 UT and 15:52:44 UT, respectively. Each $T_b$ value in the spectra represents the peak brightness temperature within the selected spatial regions at a given frequency $\nu$. The spectrum in red is obtained from the flux rope core region (source ``C''), and the spectra in blue and green are, respectively, derived from regions near the northern and southern footpoints of the flux rope (marked in Figure~\ref{fig:spec} as ``N'' and ``S''). 

The $T_b(\nu)$ spectrum of the flux rope core at the very early phase of the flare (Figure \ref{fig:spec}(b); the timing is indicated by $t_0$ in the total-power microwave/X-ray light curves in Figure \ref{fig:spec}(c)) shows a flat ``plateau'' at low frequencies with a brightness temperature approaching $\sim$30 MK. The $T_b(\nu)$ spectrum drops off precipitously above $\sim$4.4 GHz until it meets a component (dashed red line) with a shallower slope (power-law index of about $-2$). Such characteristics are consistent with thermal gyrosynchrotron and bremsstrahlung radiation from a ``superhot'' electron population \citep[e.g.,][]{1982ApJ...259..350D,2015ApJ...802..122F}: the low-frequency plateau is due to optically thick gyrosynchrotron emission from the superhot electrons at $\sim$30 MK. The sharp drop-off is due to the sudden loss of optical thickness from gyrosynchrotron at higher frequencies, where the optically thin bremsstrahlung radiation takes over. The red curve shows the best-fit result. The thermal electron temperature from the fit ($T_e \approx 30$ MK) matches the observed $T_b$ values at the low-frequency optically thick plateau. The optically thin part of the spectrum (due to breamsstrahlung) suggests an EM of $\xi = n^2_e L \approx 1.8 \times 10^{30}$ cm$^{-5}$. This microwave-derived EM value is of the same order of magnitude as that from the DEM analysis obtained by combining concurrent SDO/AIA EUV and Hinode/XRT Be-thick images using the {\tt xrt\_dem\_iterative2} method (the latter has a broad temperature response in $\sim$10--100 MK; \citealt{2007SoPh..243...63G}). Equally interestingly, the precipitous drop of $T_b$ above $\sim$4.4 GHz gives an excellent constraint for the magnetic field strength of the flux rope core. Our fit suggests $B\approx 212$ G, consistent with the value estimated from the empirical formula in \citet{1982ApJ...259..350D} (their equation 24(b)). We note that, to magnetically confine this 30 MK superhot source, a coronal magnetic field strength of at least 93 G is required (assuming a column depth of 10 Mm; see, e.g., discussions in \citealt{2010ApJ...725L.161C} and \citealt{2018ApJ...868..148L}). Our measured coronal magnetic field strength of $>$200 G is sufficiently strong to provide such magnetic confinement. We also point out that a strong coronal magnetic field strength of several hundred Gauss and above was also reported in several recent studies of the same event, although deduced at different times and locations \citep[e.g.,][]{2018ApJ...863...83G,2019ApJ...874..126K,Chen2020,Fleishman2020}.

During the slow-rise phase of the flux rope ($t_1$ in Figure~\ref{fig:spec}(c)), the spatially resolved microwave spectra from all the three sources (Figure~\ref{fig:spec}(e)) display features that are consistent with nonthermal gyrosynchrotron radiation \citep{1982ApJ...259..350D}. As shown in Figure~\ref{fig:spec}(e), the spectra have a positive and negative slope at the low- and high-frequency side (due to optically thick and thin emission, respectively), with a peak brightness temperature of $>$60 MK. We adopt the fast gyrosynchrotron codes in \citet{2010ApJ...721.1127F} to calculate the gyrosynchrotron brightness temperature spectrum from a source with model parameters (assumed to be uniform along the LOS) including magnetic field strength $B$, thermal density $n_e$, and nonthermal electron distribution with power-law index $\delta$ and a total density of $n^{\rm nth}_e$ above 10 keV. A forward fit is performed to match the calculated model spectra and the observations (see \citealt{Chen2020;Fleishman2020} for details on microwave spectral fitting). The best-fit results suggest a magnetic field strength of $B\approx 350$--480 G for all the three sources, and a nonthermal power-law index of $\delta \approx 2.5$--2.7. 

In Figure \ref{fig:spec}(f), we also show the spatially resolved light curves at 3.9 GHz derived from the three selected regions in panel (d). Each brightness temperature value on the light curves indicates the peak brightness temperature obtained within the region. The light curves of all three sources display very similar temporal evolution since the onset of the flare. They peak at about the same time ($\sim$15:54--15:55 UT), which coincides with the first impulsive peak in the total-power microwave/HXR curves (Figure \ref{fig:spec}(c)). After that, they show a rapid decay (with a half-life decay time $\tau_{1/2}$ of about one minute) prior to the primary microwave/HXR flare peak at $\sim$16:00 UT (Figure \ref{fig:spec}(c)). The very similar temporal and spectral properties of the three widely separated microwave sources (located at the flux rope core and two conjugate footpoints) indicate that they likely are magnetically connected and share a common origin.

\begin{figure*}[ht!]
\epsscale{1.1}
\plotone{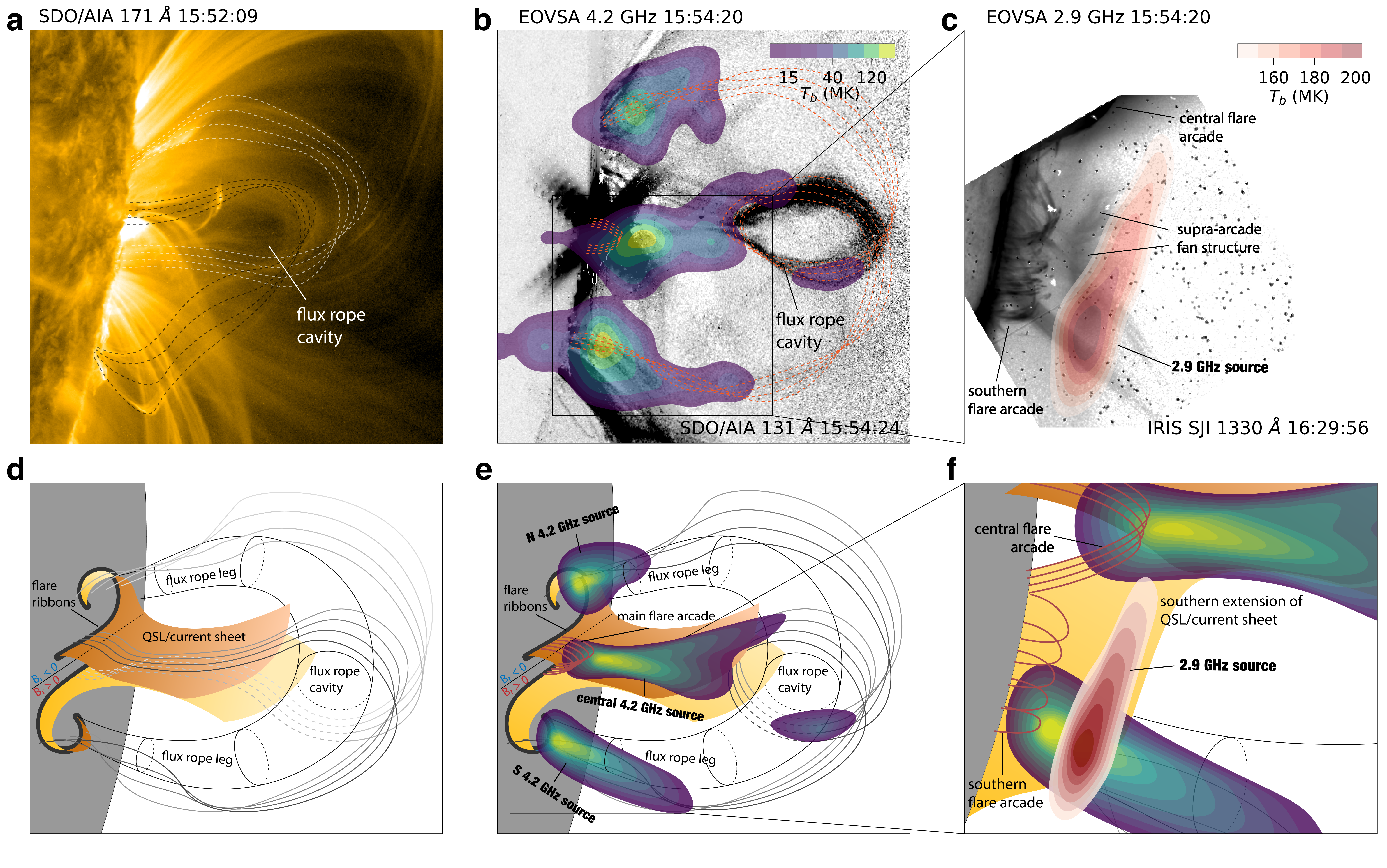}
\caption{Schematic of the microwave and EUV observations within the context of the 3D standard flare model (adapted from \citealt{2014ApJ...788...60J}). (a) SDO/AIA 171 \AA\ image just before the impulsive phase, showing loops surrounding the flux rope cavity. (b) Composite EOVSA 4.2 GHz and SDO/AIA 131 \AA\ image during the impulsive phase (same as Figure \ref{fig:fr_mw}(d)). (c) Enlarged view of the southern flare region, showing composite EOVSA 2.9 GHz image at 15:54:20 UT (red) and IRIS 1330 \AA\ slit-jaw image at 16:29:56 UT (background reverse grayscale image). (d) Schematic for the pre-impulsive phase in (a). The central orange/yellow area represents parts of the 3D reconnection current sheet and the quasi-separatrix layer (QSL) below the flux rope. The ends of the hook-shaped footprints of the 3D QSL (or flare ribbons) are located near the legs of the erupting flux rope. Thin gray curves are representative pre-reconnection field lines. (e) Schematic for the post-reconnection scenario during the impulsive phase. The reconnected field lines below the null point become part of the central flare arcade (red curves), and those above the null point join the outer rim of the flux rope body (thin gray curves). Accelerated electrons propagate along the reconnected field lines and form the central microwave source. Some can arrive near the footpoints of the flux rope and produce the northern and southern microwave side sources. (f) Schematic of the southern extension of the 3D current sheet/QSL that has a face-on viewing perspective. The 2.9 GHz source is located above the southern flare arcade that appears later on in (E)UV images. Note to better show the 3D structures, unlike the observations, the active region in the schematics of (d) and (e) is rotated slightly inside the west limb. \label{fig:cartoon}}
\end{figure*}

\section{Discussion} \label{sec:discussion}
Our observational results are broadly consistent with the magnetic topology and the associated energy release scenario suggested in the 3D standard model for eruptive flares \citep{2012A&A...543A.110A,2013A&A...549A..66A,2013A&A...555A..77J}. Figure \ref{fig:cartoon}(a) shows a schematic picture of the model adapted from \citet{2014ApJ...788...60J,2015SoPh..290.3425J}. Prior to the eruption, a preexisting magnetic flux rope, visible as a reverse S-shaped dark filament with an overall north--south orientation, is located at the PIL separating the two main sunspot groups with opposite magnetic polarity. Such a reverse S-shape of the filament implies a negative magnetic helicity of the flux rope \citep[e.g.,][]{2012A&A...543A.110A}, which is typical for the northern hemisphere \citep{1995ApJ...440L.109P}. The center of the filament has an east--west orientation along the main PIL (Figure \ref{fig:fila}(b)). When the AR rotated to the limb on September 10, this central portion of the filament became nearly aligned with the LOS perpendicular to the plane of the sky (Figure \ref{fig:fila}(d)). Therefore, when the flux rope erupted, its central portion developed into a dark cavity with an LOS-aligned axis. The two legs of the flux rope (which are displayed as the two ``hooks'' of the filament northward and southward from its center) exhibit in EUV/SXR as hot channel strands connecting the central cavity from both the northern and southern side. Reconnection between the progenitor field lines (represented as thin gray curves in Figures \ref{fig:cartoon}(a) and (d)) surrounding the flux rope form an extended 3D current sheet immediately beneath it (orange/yellow plane in Figure \ref{fig:cartoon}(d)). The central portion of the current sheet, therefore, has an edge-on viewing perspective similar to the central flux rope cavity.
Intense heating of the plasma surrounding the central current sheet and the newly reconnected field lines give rise to the thin plasma-sheet-like structure and the bright outer rim of the flux rope cavity seen in hot EUV/SXR passbands (see, e.g., 3D numerical modeling results in \citealt{2019ApJ...887..103R}). \citet{Chen2020} derived the magnetic field variation along the plasma sheet feature based on EOVSA imaging spectroscopy data, which matched very well with theoretical predictions in the standard model that includes an extended current sheet. Such plasma heating may also explain the superhot ($\sim$30 MK) flux rope core observed early on in microwaves (see Figures \ref{fig:spec}(a) and (b)).

In contrast to the orientation of the central portion of the 3D current sheet along the LOS, the northern and southern extension of the current sheet had a nearly face-on viewing perspective (orange shaded area in Figure \ref{fig:cartoon}(f)). Such a face-on current sheet is strongly implicated by the presence of an elongated looptop microwave source at 2.9 GHz that displays the same north--south, parallel-to-the-limb orientation, shown as red filled contours in Figure \ref{fig:cartoon}(c). The 2.9 GHz source is located at or slightly above the location of a hot supra-arcade fan (SAF) structure, which appeared a few minutes later (after $\sim$16:00 UT) in AIA hot passbands (94, 131, and 193 \AA) and IRIS 1330 \AA\ slit-jaw images (which have a contribution from the \ion{Fe}{21} line sensitive to $\sim$10 MK plasma in addition to the \ion{C}{2} transition region line; \citealt{2019MNRAS.489.3183C}). At later times, a series of post-flare arcades become visible in cooler AIA passbands and IRIS 1330 \AA\ slit-jaw images below the SAF structure, first southward and later also northward of the main flaring region, which extend in both directions along the west limb (see, e.g., Figure \ref{fig:fila}(f)). Such SAF structures have been interpreted as heated plasma near the base of a large-scale current sheet with a face-on viewing geometry \citep[see, e.g.,][and references therein]{2017ApJ...836...55R,2019MNRAS.489.3183C}. All these signatures strongly support the presence of a face-on, north--south extension of the 3D current sheet trailing the eruption of the flux rope/filament that likewise has a north--south-oriented component.

As discussed in \citet{2012A&A...543A.110A}, during the early eruption phase, the pre-reconnection field lines are highly sheared. While the inner ends of these field lines are rooted close to the PIL and reconnect at the central current sheet, their two other ends are rooted near the legs of the flux rope (thin gray curves in Figure \ref{fig:cartoon}(a) and corresponding observations in SDO/AIA 171 \AA\ image shown in Figure \ref{fig:cartoon}(b)). After the reconnection, the newly reconnected field lines above the primary reconnection X point join the flux rope and add to its magnetic flux (thin gray curves in Figure \ref{fig:cartoon}(c)). 
The other set of the reconnected field lines below the X point, in turn, form the main flare arcade (red curves in Figure \ref{fig:cartoon}(c)). The majority of nonthermal electrons are presumably accelerated at or below the central current sheet and escape along the newly reconnected field lines \citep[see observations by, e.g.,][] {2003ApJ...596L.251S,2018ApJ...866...62C,Chen2020}. The north--south extension of the 3D current sheet may also be responsible for energizing electrons to nonthermal energies, as evidenced by the 2.9 GHz source, although these electrons may be relatively smaller in number compared to the central flaring site. The electrons accelerated in the central current sheet region traveling upward can gain access to the outer shell of the flux rope and propagate back to the surface near the flux rope footpoints. Hence, microwave sources can be detected wherever these nonthermal electrons accumulate sufficient density to produce emission (via gyrosynchrotron radiation) with a brightness temperature high enough to be distinguished against the background. 

Such microwave sources are vividly shown in EOVSA images at low frequencies ($<$4 GHz). They appear at various sites that include the central current sheet region, with an extension upward encompassing the flux rope cavity, as well as the flux rope footpoints and legs (Figures \ref{fig:cartoon}(c) and (d)). Excluding the top of the central flare arcade where the microwave emission is the most prominent (which may be the site of the primary electron acceleration; \citealt{2010ApJ...714.1108K,Chen2020,Fleishman2020}), the brightest microwave emission occurs at the bottom of the flux rope cavity and the two far ends of the flux rope while emission elsewhere in the flux rope is faint or absent. We attribute this phenomenon to the spatial variation of the magnetic field strength along the flux rope: the relatively strong magnetic field strength in these microwave-bright regions allows the microwave emission to peak in the 3--4 GHz range (see, spectra in Figure \ref{fig:spec}(e)). A lower magnetic field strength would effectively shift the entire spectrum to the lower frequencies (see, e.g., Movie S2 of \citealt{Fleishman2020} for illustration). In this case, the microwave brightness at 3--4 GHz and above would be dominated by weak optically thin emission and become much fainter. A similarly weak magnetic field may explain why the microwave source at the location of the north--south extension of the postulated 3D current sheet is only observed at our lowest observing frequency of 2.9 GHz (see, Figures \ref{fig:spec}(a) and (d)).

The two microwave side sources near the flux rope footpoints quickly diminish after the first microwave/HXR impulsive peak at $\sim$15:54 UT. This feature may be related to the strong-to-weak shear evolution of the reconnecting field lines as the flare progresses \citep{2012A&A...543A.110A}: early on in the flare when the flux rope remains low in the corona, the remote ends of the strongly inclined reconnecting field lines are anchored near the flux rope footpoints, allowing the accelerated electrons to access the footpoint region easily (left panels of Figure 5 in \citealt{2012A&A...543A.110A}). Hence the double microwave side sources near the flux rope footpoints are observed. As the flux rope lifts off to greater heights (for example, by 16:00 UT, the flux rope/CME has already reached $>1R_{\odot}$ above the surface; \citealt{2018ApJ...868..107V}), the system has evolved to higher pre-reconnection field lines that have less shear---eventually these become nearly coplanar in the plane of the sky, perpendicular to the primary PIL. Their roots are located away from the flux rope footpoints close to the PIL (right panels of Figure 5 in \citealt{2012A&A...543A.110A}). Therefore, the flare geometry almost returns to the 2D CSHKP scenario, in which the accelerated electrons cannot find an easy path to reach the regions near the flux rope footpoints. Hence a rapid decay of the microwave side sources is observed.

We note that nonthermal side sources near the flux rope footpoints were previously depicted in the celebrated semi-3D standard flare cartoon by \citet{1995ApJ...451L..83S}. However, to our knowledge, reports of these remote side sources have been elusive. In this event, we also found no HXR counterpart at the same location in the RHESSI data. One possible reason is that they are relatively weak and short-lived compared to the central main source, and are thereby difficult to detect with RHESSI's limited dynamic range and sensitivity. Perhaps more importantly, under our scenario, the electrons responsible for these sources need to propagate for a much greater distance (from the acceleration site to flux rope body and back to flux rope footpoints) than their counterparts at the footpoints of the central flare arcade. The HXR-emitting electrons at tens of kiloelectron volts traveling to the flux rope footpoints suffer significant loss due to, e.g., Coulomb collisions. In contrast, the microwave-emitting, more energetic electrons (hundreds of kiloelectron volts to above million electron volts) are more likely to reach the flux rope footpoints without significant loss, as the Coulomb stopping column is proportional to the square of the electron energy (i.e., $N^{\rm cc} = n_e L \approx 10^{17} (E/{\rm keV})^2$ cm$^{-2}$; \citealt{1988psf..book.....T}).

Our study exemplifies the potential of microwave observations in detecting and diagnosing flux ropes (and other magnetic structures) illuminated by flare-accelerated electrons. Since its completion in 2017, EOVSA has provided a new view for flares thanks to its unprecedented microwave imaging spectroscopy capability. Yet EOVSA's image fidelity and dynamic range are inevitably limited by its small number of antennas (13) available for imaging. Future facilities, such as the Frequency Agile Solar Radiotelescope concept (FASR; \citealt{2019astro2020U..56B}), will provide high-fidelity broadband imaging spectroscopy with orders-of-magnitude-improved dynamic range (10$^4$:1 for FASR vs. 10--100:1 for EOVSA) thanks to its larger number of antennas (64 in 2--20 GHz and 48 in 0.2--2 GHz) and much denser u-v coverage (14--26 times more baselines than EOVSA). Observations from these facilities will open up a new window for fully exploring the magnetic energy release, electron acceleration, and electron transport processes throughout the flare region from low- to mid-corona.


\acknowledgments

EOVSA operation is supported by NSF grant AST-1910354. B.C., S.Y., and D.G. are supported by NSF grants AGS-1654382 and AGS-1723436 to NJIT. K.R. is supported by NSF grant AGS-1923365 to SAO. The work is supported partly by NASA DRIVE Science Center grant 80NSSC20K0627, and NASA grant 80NSSC18K1128 to NJIT. We are grateful to Drs. Gregory Fleishman and Alexey Kuznetsov for making their fast gyrosynchrotron codes publicly available. We thank the Royal Observatory of Belgium for providing the USET H-$\alpha$ data, and the SDO/AIA and SDO/HMI teams for providing the (E)UV and vector magnetogram data. Hinode is a Japanese mission developed and launched by ISAS/JAXA, with NAOJ as domestic partner and NASA and STFC (UK) as international partners. It is operated by these agencies in cooperation with ESA and NSC (Norway).

\vspace{0mm}
\facilities{OVRO:SA, SDO, Hinode}

\software{CASA \citep{2007ASPC..376..127M},
          Astropy \citep{2018AJ....156..123A}, 
          SunPy \citep{sunpy_community2020}
          }
          
\appendix

\vspace{-5mm}

\section{EOVSA Data Processing}\label{ap:imaging}
EOVSA observed the event in 2.5--18 GHz in 134 frequencies spread over 31 equally spaced spectral windows (``SPW''). Except for the lowest-frequency window centered at 2.92 GHz (``SPW 0''), all the other 30 SPWs (SPWs 1--30) were calibrated in phase against an unresolved celestial source. Self-calibration is performed for each of the SPWs 1--30 using a time near the peak of the impulsive phase (at $\sim$15:59 UT) when the microwave emission displays a simple single-source geometry at all frequencies. For SPW 0, as there is no available phase calibration from a celestial source, we use the nearby window (SPW 1 centered at 3.4 GHz) as the model for self-calibration. The self-calibration solutions for all the SPWs are applied uniformly to the visibility data from 15:46 UT to 16:06 UT, which are subsequently used for further time-dependent self-calibration and synthesis imaging. The absolute flux calibration is done by scaling the total flux of the images at all frequencies to the concurrent total-power data near the flare peak. 

We have carefully assessed the resulting synthesized images in the spatial, time, and frequency domains to ensure that the bright microwave sources within the time of interest are not affected by side lobes of the synthesized beam. We find that images at SPWs 2 and 3 (centered at 3.9 GHz and 4.4 GHz, respectively) best show counterparts of the EUV/SXR hot channel feature during its eruption. In order to better reveal the detailed morphology of the evolving flux rope in microwaves, at each given time integration, we adopt a multifrequency synthesis (MFS) image deconvolution technique \citep{2011A&A...532A..71R}, available in CASA's ``tclean'' task, to combine all the available 11 spectral channels in SPWs 2 and 3 (which cover the frequency range from 3.85 GHz to 4.50 GHz) and form a single image. This practice effectively increases the ``uv-coverage''---sampling of the visibility function of the sky brightness distribution at a discrete set of spatial frequencies $D_{ij}/\lambda_{k}$ (where $D_{ij}$ is the baseline length between an antenna pair $i$ and $j$, and $\lambda=c/\nu_{k}$ is the wavelength of a spectral channel $k$), thereby improves the image dynamic range and fidelity. 

To correct for time-dependent, small variations of antenna gains, we perform further self-calibrations for each of the 4 s integrated microwave visibility data. We use the auto-masking technique incorporated in the ``tclean'' task of CASA (Common Astronomy Software Applications v5.4; \citealt{2007ASPC..376..127M}) to automatically identify source regions in the microwave images during the CLEAN process. In each self-calibration cycle, the CLEAN components within the auto-masked region were used as the model to derive phase and/or amplitude solutions to be applied back to the visibility data for correction. During each self-calibration cycle, the gain solutions are relatively small, i.e., a few degrees in phase and $\lesssim$10\% in amplitude, and the general source morphology stays unchanged. However, this practice effectively reduces artifacts in the images due to imperfect calibration solutions. The image dynamic range (defined as the ratio of the maximum brightness in the image to the root mean square of the brightness in an empty region without any source $D = {\rm max}(T_b):\sigma(T_b)$) is improved by $\sim$22\%--149\% (median 34\%) after the time-dependent self-calibration. The nominal FWHM angular resolution of the microwave spectral images is $113''.7/\nu_{\rm GHz}\times53''.0/\nu_{\rm GHz}$. In this study, all the single-SPW CLEAN images are restored using a circular restoring beam with a size of $73''.0/\nu_{\rm GHz}$, while the size is fixed at 5$''$ above 14.5 GHz (note that here we used a slightly smaller restoring beam than \citealt{2018ApJ...863...83G}). A circular restoring beam with an FWHM of $20''$ is used for restoring the 3.85--4.50 GHz frequency-synthesis images using SPWs 2 and 3. 

\bibliographystyle{aasjournal}

\end{document}